\documentclass[12pt,letter]{article}
\usepackage{graphicx}
\usepackage{amsmath}
\usepackage{amssymb}
\usepackage{cancel}
\usepackage[left=2cm,right=2cm,top=3cm,bottom=2cm]{geometry}
\usepackage{setspace}
\usepackage[super,sort&compress,comma]{natbib} 
\usepackage{bm}
\usepackage[T1]{fontenc}
\usepackage{microtype}
\usepackage{subfigure}
\usepackage{chemformula}
\usepackage{tabularx}
\usepackage{booktabs}
\usepackage{multirow}
\usepackage{mathtools}
\usepackage{bm}
\usepackage{esvect}

\begin{document}
\title{\textbf{NewtonNet: A Newtonian message passing network for deep learning of interatomic potentials and forces}}
\date{}
\author{Mojtaba Haghighatlari$^{1}$, Jie Li$^{1}$, Xingyi Guan$^{1,5}$, Oufan Zhang$^{1}$,
\\
Akshaya Das$^{1}$,Christopher J. Stein$^{1,5-6}$,Farnaz Heidar-Zadeh$^{1-3}$,Meili Liu$^{1,4}$,
\\Martin Head-Gordon$^{1,5}$,Luke Bertels$^{1}$, Hongxia Hao$^{1,5}$,Itai Leven$^{1,5}$,
\\
Teresa Head-Gordon$^{1,5,7}$}
\maketitle
\noindent
$^1$Kenneth S. Pitzer Theory Center and Department of Chemistry, University of California, Berkeley, CA, USA\\
$^2$Center for Molecular Modeling (CMM), Ghent University, B-9052 Ghent, Belgium\\
$^3$Department of Chemistry, Queen's University, Kingston, Ontario K7L 3N6, Canada\\
$^4$Department of Chemistry, Beijing Normal University, Beijing, China\\
$^5$Chemical Sciences Division, Lawrence Berkeley National Laboratory, Berkeley, CA, USA\\
$^6$ Theoretical Physics and Center for Nanointegration Duisburg-Essen (CENIDE), University of Duisburg-Essen, 47048 Duisburg, Germany\\
$^7$Departments of Bioengineering and Chemical and Biomolecular Engineering, University of California, Berkeley, CA, USA\\
\begin{center}
corresponding author: thg@berkeley.edu
\end{center}

\begin{abstract}
\noindent
We report a new deep learning message passing network that takes inspiration from Newton's equations of motion to learn interatomic potentials and forces. With the advantage of directional information from trainable latent force vectors, and physics-infused operators that are inspired by the Newtonian physics, the entire model remains rotationally equivariant, and many-body interactions are inferred by more interpretable physical features. We test NewtonNet on the prediction of several reactive and non-reactive high quality \textit{ab initio} data sets including single small molecule dynamics, a large set of chemically diverse molecules, and methane and hydrogen combustion reactions, achieving  state-of-the-art test performance on energies and forces with far greater data and computational efficiency than other deep learning models. 
\end{abstract}
\newpage
\section*{\fontsize{12}{12}\selectfont INTRODUCTION}
\label{sec:intro}
\noindent
The application of machine learning models to predict \textit{ab initio} potential energies for bulk silicon by Behler and Parrinello\cite{Behler2007} less than 15 years ago has inspired the modern cheminformatics era. Their idea of approximating the total potential energy as the sum of atomic energies paved the way for the hierarchical decomposition of structural and physical features of molecules in order to provide atom-centric feature representations, and which more easily addresses different permutations and/or number of atoms in molecules. With such strategies, machine learning (ML) models have provided good to accurate predictions of molecular energy and atomic forces\cite{Cendagorta2020, Unke2021review}, with emerging abilities to drive Newton's equations of motion in molecular simulation.\cite{schutt2018schnetpack, Wang2018b, zeng2020complex} These ML endeavors have typically required prodigious quantities of data, as exemplified by the 57,000 small CHNO-containing molecules perturbed into more than 22 million different configurations with energies evaluated with Density Functional Theory (DFT)\cite{Parr1994} data in the so-called "ANI-1" data sets.\cite{smith2017ani,Devereux2020,Smith2020}

In earlier artificial neural networks (ANNs) and kernel methods\cite{Haghighatlari2020review} used to predict energies as well as vectorial forces, the atomic environments are modeled using symmetry functions that rely on two-body and higher order correlated features, e.g., distances, angles, dihedrals, etc., for any central atom.\cite{Behler2007,smith2017ani,Wang2018a} Despite the rising computational cost of incorporating 3-body representations and above, these hand-crafted many-body representations have shown significant improvements in predicting interatomic potentials and directional forces\cite{Faber2018, Christensen2020}. Alternatively, message passing neural networks\cite{Gilmer2017} (MPNN) replace the hand-crafted features of the distances and angles with trainable operators that only rely on the atomic numbers and positions, such that the learned latent space of MPNNs have an added advantage in chemical accuracy compared to explicit symmetry functions. A major contributing MPNN method for 3D structures is SchNet\cite{schutt2018schnetpack}, which takes advantage of a continuous filter layer that facilitates the convolution of decomposed interatomic distances with atomic attributes. Related methods have subsequently built on this success in incorporating additional features to describe atomic environments. For example, PhysNet\cite{Unke2019a} adds prior knowledge about the long-range electrostatics in energy predictions, and DimeNet \cite{Klicpera2020} takes advantage of angular information and more stable basis functions based on Bessel functions. 

But in the standard MPNN, the representation is usually reduced to transformationally identical features, for example quantities that are invariant to translation and permutation. However, by only representing the atomic features to be invariant to the transformations of the system, Pozdnyakov et. al.\cite{Pozdnyakov2020} has shown that all such models, including MPNNs, fail in distinguishing systems even if represented by up to 4-body information. Because we aim to predict not only energies but force vectors, and given the fact that vectorial features can be affected by transformation $T$ of input structure $\bm{x} \in X$, we need to ensure the output of each operator also will reflect such transformation equivalently when needed. More specifically, rotational transformations (such as through angular displacements $\phi$) are one of the biggest challenges in the modeling of 3D objects, illustrated in learning a global orientation of structures for MD trajectories with many molecules, that is very difficult or infeasible. Mathematically speaking, the function $\phi: X \rightarrow Y$ is equivariant to a transformation group $g$, if 
\begin{align}
    \phi(T_g(\bm{x})) = T_g^{'}(\phi(\bm{x})),
\end{align}
where $T_g$ and $T_g^{'}$ are equivalent transformations in that abstract group acting on the input and output space, respectively\cite{Satorras2021}. 

Only very recently have neural networks been developed that are equivariant to the transformations in Euclidean space, and are emerging as state-of-the-art ML methods in predictive performance when evaluated on a variety of tasks that are fast superseding invariant-only models. Furthermore, equivariant models are found to greatly reduce the need for excessively large quantities of reference data, ushering in a new era for machine learning on the highest quality but also the most expensive of \textit{ab initio} data. For instance, a group of machine learning models have introduced multipole expansions such as used in NequIP\cite{Kondor18,thomas18,Anderson2019,batzner21}, or are designed to take advantage of precomputed features and/or higher-order tensors using molecular orbitals\cite{Qiao2021, Glick2021}.  PaiNN\cite{Schutt2021} is a MPNN model that satisfies equivariance, but the mathematical operations do not follow a physically interpretable procedure. In spite of the added advantage of infusing extra physical knowledge into machine learning models, the computational cost of spherical harmonics and availability/versatility of pre-computed features, or lack of physical interpretability, can be limiting. In particular, an equivariant model should be equipped with appropriate strategies to decrease the rank of equivariant features needed throughout the architecture of the model to be computationally viable while also retaining test set accuracy.

\begin{figure*}[!htb]
\center
\includegraphics[width=0.95\textwidth]{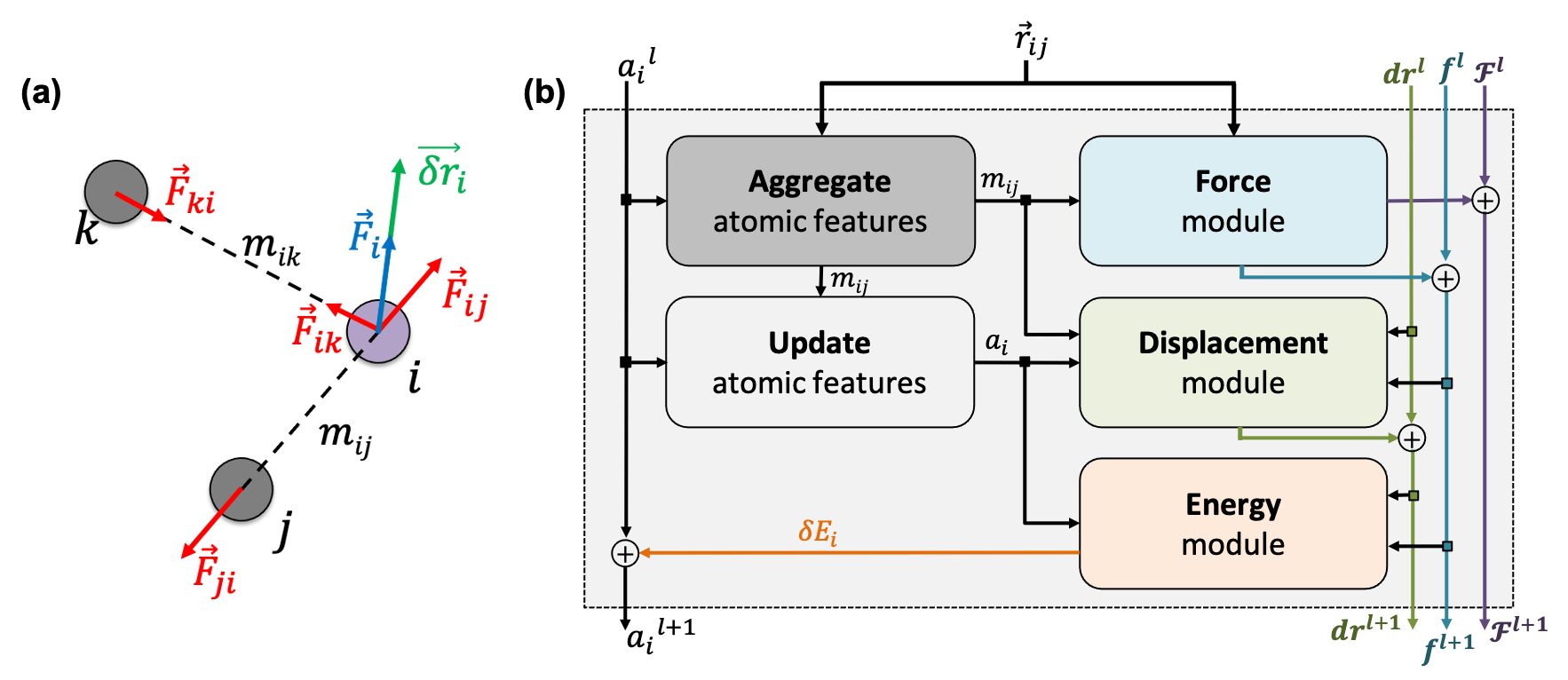}
\caption{(a) Newton's laws for the force and displacement calculations for atom $i$ with respect to its neighbors. (b) Schematic view of the NewtonNet message passing layer. At each layer four separate components are updated: atomic feature arrays $a_i$, latent force vectors $\bm{\mathcal{F}}$, and force and displacement feature vectors ($\bm{f}$ and $\bm{dr}$).  }
\label{fig:latent_force}
\end{figure*}

In this work we introduce a geometric MPNN \cite{Monti2017} based on Newton's equations of motion that achieves equivariance with respect to physically relevant permutations. NewtonNet improves the capacity of structural information in the ML model by creating latent force vectors based on the Newton's third law (Fig. 1a). The force direction helps to describe the influence of neighboring atoms on the central atom based on their directional positions of atoms in the 3D space with respect to each other. Since we now introduce vector features as one of the attributes of atoms, we thereby enforce the model to remain equivariant to the rotations in the 3D coordinate space and preserve this feature throughout the network. By infusing more physical priors into the network architecture, NewtonNet realizes a computational cost that is more favorable, and enabling modeling of reactive and non-reactive chemistry with superior performance to other equivariant models, and doing so with reductions down to only 1-10\% of the original training data set sizes needed for invariant-only ML models.

\section*{\fontsize{12}{12}\selectfont THEORY}
\noindent
Given a molecular graph $\mathcal{G}$ with atomic features $a_i \in \mathbb{R}^{\mathrm{nf}}$ (where nf is the number of features) and interatomic attributes ${e_{ij}} \in \mathbb{R}^{\mathrm{b}}$, a message passing layer can be defined as\cite{Gilmer2017}:

\begin{align}
    {m}_{i j} &=M_{l}\left({a}_{i}^{l}, {a}_{j}^{l}, {e_{i j}}\right)
\end{align}
\begin{align}
   {m}_{i} &=\sum_{j \in \mathcal{N}(i)} {m}_{i j}
    \label{eqn:aggregate}
\end{align}
\begin{align}
    {a}_{i}^{t+1} &=U_{l}\left({a}_{i}^{l}, {m}_{i}\right)
\end{align}
\noindent
where $M_l$ is the message function and $U_l$ is called the update function, and the sub-/super-script $l$ accounts for the number of times the layer operates iteratively. A combination of explicit differentiable functions and operators with trainable parameters are the common choice for $M_l$ and $U_l$. 

NewtonNet considers a molecular graph defined by atomic numbers $Z_{i} \in \mathbb{R}^{\mathrm{1}}$ and relative position vectors ${\vv{r_{ij}}=\vv{r_{j}} - \vv{r_{i}}} \in \mathbb{R}^3$, as input and applying operations that are inspired by Newton's equations of motion to create features arrays ${a}_{i}$ $\in$ $\mathbb{R}^{\mathrm{nf}}$, to represent each atom in its immediate environment, while remaining invariant to the rotations of the input configuration. To do this, NewtonNet takes advantage of multiple layers of message passing which are rotationally equivariant, in which each layer consists of multiple modules that include operators to construct force and displacement feature vectors, which are contracted to the feature arrays via the energy calculator module (Fig. 1b). 
We emphasize the critical role of projecting equivariant feature vectors to invariant arrays since one goal of the model is to predict potential energies, which are invariant to the rotations of atomic configurations. Consequently, we avoid any many-to-one mapping of rotationally equivariant features to invariant energies. The core idea behind the iterative message passing of the atomic environments is to update the feature array ${a}_{i}^{t}$ that represent each atom in its immediate environment. This level of feature representation remains equivariant to the rotations of the initial configuration in NewtonNet through the following operations: atomic feature aggregator, and energy, force and displacement modules. We provide the proof of equivariance of the NewtonNet model in Appendix A. 

\textbf{Atomic feature aggregator.} This is a standard message passing module that is invariant to the rotation, but with a notable difference to early MPNNs in use of a symmetric message function (i.e., $m_{ij}=m_{ji}$) between atom pairs, and is used in all the equivariant modules to account for the interatomic interactions (Fig. 1b). We initialize the atomic features based on trainable embedding ${g}: \mathbb{R}^{\mathrm{1}} \rightarrow \mathbb{R}^\mathrm{nf}$ of atomic numbers $Z_i$, i.e., ${a}_{i}^{0} = g(Z_i)$. We next use the edge function ${e}: \mathbb{R}^{\mathrm{3}} \rightarrow \mathbb{R}^\mathrm{nb}$ to represent the interatomic distances using radial Bessel functions as introduced by Klicpera et al.\cite{Klicpera2020}

\begin{align}
    e(\vv{r_{i j}})=\sqrt{\frac{2}{r_{c}}} \frac{\sin (\frac{n \pi}{r_{c}} \lVert \vv{r_{i j}} \rVert )}{\lVert \vv{r_{i j}} \rVert}
\end{align}
\noindent
where $r_c$ is the cutoff radius and $\lVert \vv{r_{i j}} \rVert$ returns the interatomic distance between any atom $i$ and $j$.
We follow Schutt et al.\cite{Schutt2021} in using a self-interaction linear layer $\phi_{rbf}: \mathbb{R}^{\mathrm{nb}} \rightarrow \mathbb{R}^\mathrm{nf}$ to combine the output of radial basis functions with each other. This operation is followed by using an envelop function to implement a continuous radial cutoff around each atom through use of a polynomial function $e_{cut}$ introduced by Klicpera et al.\cite{Klicpera2020} with the polynomial degree $p=7$. Thus, the edge operation ${\phi_e}: \mathbb{R}^{\mathrm{3}} \rightarrow \mathbb{R}^\mathrm{nf}$ is defined as a trainable transformation of relative atom position vectors in the cutoff radius ${r}_{c}$,

\begin{align}
    \phi_{e}\left(\vv{r_{i j}}\right)=\phi_{\operatorname{rbf}}\left(e\left(\vv{r_{i j}}\right)\right) e_{c u t}\left(r_{c},\left\|\vv{r_{i j}}\right\|\right).
\end{align}
\noindent
Hence the output of $\phi_{e}$ is rotationally invariant as it only depends on the interatomic distances.

A symmetric message $m_{ij}$ is passed between atom $i$ and atom $j$ that are the same in both directions, and introduced by element-wise product between all feature arrays involved in any two-body interaction,
\begin{align}
    {m}_{i j} = \phi_{a}({a}_{i}^{l}) \left. \phi_{a}({a}_{j}^{l})\right. \phi_{e}(\vv{r_{i j}}),
\end{align}
\noindent
where $\phi_{a}: \mathbb{R}^{\mathrm{nf}} \rightarrow \mathbb{R}^\mathrm{nf}$ indicates a trainable and differentiable multilayer perceptron\cite{Haghighatlari2020review}.
Note that the $\phi_{a}$ is the same function applied to all atoms. Thus, due to weight sharing and multiplication of output features of both atoms of the two-body interaction, the ${m}_{i j}$ remain symmetric at each layer of message passing. To complete the feature array aggregator, we use the equation \ref{eqn:aggregate} to simply sum all messages received by central atom $i$ from its neighbors $\mathcal{N}(i)$. Finally, we update the atomic features at each layer using the sum of received messages,
\begin{align}
    {a}_{i}^{l+1} = {a}_{i}^{l} + \sum_{j \in \mathcal{N}(i)} {m}_{ij}.
\end{align}

\textbf{Force module}. We take advantage of directional information in the force calculator module in which we can estimate the symmetric force magnitude as a function of ${m}_{i j}$, i.e., $\lVert \vv{F}_{ij}\rVert = \phi_{F}({m}_{i j})$. The product of the force magnitude by unit distance vectors $\hat{r}_{i j} = \vv{r}_{i j}/\lVert \vv{r}_{i j} \rVert$ gives us antisymmetric interatomic forces that then obey Newton's third law (note that $\vv{r}_{i j} = -\vv{r}_{j i}$),
\begin{align}
    \label{eq:fij}
    \vv{F}_{ij}^{l} = \left.\phi_{F}({m}_{i j}\right.) \hat{r}_{i j}
\end{align}
where $\phi_{F}$ is a differentiable learned function. The total force $\vv{F}_{i}^{l}$ at each message passing layer $l$ on atom $i$ is the sum of all the forces from the neighboring atoms $j$ in the atomic environment,  
\begin{align}
    \label{eq:fi}
    \vv{F}_{i}^{l} = \sum_{j \in \mathcal{N}(i)} \vv{F}_{ij}^{l},
\end{align}
and update the latent force vectors at each layer,
\begin{align}
    \bm{\mathcal{F}}_{i}^{l+1} = \bm{\mathcal{F}}_{i}^{l} + \vv{F}_{i}^{l}.
\end{align}
\noindent
We can ultimately use the latent force vector from the last layer $L$, $\bm{\mathcal{F}}_{i}^{L}$ in the loss function to ensure this latent space truly mimics the underlying physical rules. 

To complete the force calculator module, we borrow the idea of a continuous filter from Schutt et al.\cite{schutt2018schnetpack} to decompose and scale latent force vectors along each dimension using another learned function $\phi_{f}$. 
\begin{align}
    \bm{\Delta f}_{i} = \sum_{j \in \mathcal{N}(i)} \left.\phi_{f}({m}_{ij}\right.) \vv{F}_{ij}^{l}.
\end{align}
This way we can featurize the vector field to avoid too much abstraction in the structural information that they carry, and as a result the constructed latent interatomic forces are decomposed by rotationally invariant features along each dimension, which we term feature vectors. Following the message passing strategy, we update the force feature vectors with $\bm{\Delta f}_{i}$ after each layer
\begin{align}
    \bm{f}_{i}^{l+1} = \bm{f}_{i}^{l} + \bm{\Delta f}_{i}.
\end{align}

\textbf{Displacement module}. Inspired by Newton's second law that forces update displacements, we approximate the displacement vectors using a learned function $\phi_{r}$ that acts on the current state of each atom presented by its atomic features ${a}_{i}^{l+1}$,
\begin{align}
    \bm{\delta r}_{i} = \phi_{r}({a}_{i}^{l+1}) \bm{{f}_{i}^{l+1}}.
\end{align}
We finally update the displacement feature vectors by $\bm{\delta r}_{i}$ and a weighted sum of all the atomic displacements from the previous layer. The weights are estimated based on a trainable function of messages ($\phi_{r}^{'}$) between atoms,
\begin{align}
\label{eq:dr_ext}
    \bm{dr}_{i}^{l+1} = \sum_{j \in \mathcal{N}(i)} \phi_{r}^{'} \left({m}_{i j}\right) \bm{dr}_{j}^{l} + \bm{\delta r}_{i}. 
\end{align}
The weight component in this step works like an attention mechanism to concentrate on the two-body interactions that cause maximum movement in the atoms. Since forces at $l=0$ are zero (i.e., $\bm{{f}_{i}^{0}} = \bm{0}$), the displacements are also initialized with zero values, i.e., $\bm{{dr}_{i}^{0}} = \bm{0}$.

\textbf{Energy module}. The last module contracts the directional information to the rotationally invariant atomic features. Since we developed the previous force and displacement modules based on the Newton's equations of motion, one immediate idea is to approximate the potential energy change for each atom using $f_i^l$ and $\delta r_i^l$, considering that $\bm{f_i^l} \approx - \delta U / \bm{\delta r_i^l}$. Thus, we find energy change for each atom by
\begin{align}
\label{eq:du}
    {\delta U}_{i} = - \phi_{u}({a}_{i}^{l+1}) \left (\bm{f}_{i}^{l+1} \cdot \bm{dr}_{i}^{l+1}\right),
\end{align}
where $\phi_{u}$ is another differentiable learned function that operate on the atomic features and predicts the energy coefficient for each atom. The dot product of two feature vectors contracts the features along each dimension to a single feature array. We finally update the atomic features once again using the contracted directional information presented as atomic potential energy change,
\begin{align}
    {a}_{i}^{l+1} = {a}_{i}^{l+1} + {\delta U}_{i}.
\end{align}
This approach is both physically and mathematically consistent with the rotational equivariance operations and the goals of our model development. Physically, the energy change is the meaningful addition to the atomic feature arrays as they are used to predict the atomic energies eventually. Mathematically, the dot product of two feature vectors contracts the rotationally equivariant features to invariant features similar to euclidean distance that we used in the atomic feature aggregator module. 

\section*{\fontsize{12}{12}\selectfont RESULTS}
\noindent
\textbf{Single Small Molecules}. We first evaluate the performance of NewtonNet on the data generated from molecular dynamics trajectories using Density Functional Theory (DFT)\cite{Parr1994} for 9 small organic molecules from the MD17 benchmark. \cite{Chmiela2017a, Chmiela2018} Despite reported outliers in the calculated energies associated with this data\cite{Christensen2020}, we still use the original version of MD17 to predict energy and forces for each "dedicated" molecule separately. For training NewtonNet, we select a challenging data size of 950 for training, 50 for validation, and remaining data for test. This data split is more ambitious than that used by kernel methods such as sGDML\cite{Chmiela2019} and FCHL19\cite{Christensen2020}, and is supported by other emerging machine learning models that utilize equivariant operators, e.g., NequIP\cite{batzner21} and PaiNN\cite{Schutt2021} that train on a  (relatively) few number of samples. Table \ref{tab:md17} shows the performance of NewtonNet for both energy and forces on the hold-out test set, illustrating that it can outperform invariant deep learning models (e.g., SchNet\cite{schutt2018schnetpack}, PhysNet\cite{Unke2019a}, and DimeNet\cite{Klicpera2020}) and even in some cases state-of-the-art equivariant models such as NequIP and PaiNN. Furthermore, NewtonNet remains computationally efficient and scalable relative to newer methods that incorporate higher order tensors in the equivariant operators and/or are trained on a revised version of MD17 data set\cite{batzner21, Qiao2021}, while still retaining chemical accuracy (< 0.5 kcal/mol).

\begin{table*}[ht]
\centering
\footnotesize   
\caption{The performance of models in terms of mean absolute error (MAE) for the prediction of energies (kcal/mol) and forces (kcal/mol/\AA) of molecules in the MD17 data sets. We report results by averaging over four random splits of the data to define standard deviations. Best results in the standard deviation range are marked in bold.}

\begin{tabular}{llccc|cc|ccc}
        \hline\hline
\multicolumn{2}{l}{} & SchNet & PhysNet & DimeNet & FCHL19 & sGDML & NequIP & PaiNN & NewtonNet \\ \hline
\multirow{2}{*}{\textbf{Aspirin}} & energy & 0.370  & 0.230   & 0.204   & 0.182 & 0.19 & - & \textbf{0.159} & \textbf{0.168}$\pm$ 0.019 \\
& forces  & 1.35 & 0.605 & 0.499 & 0.478 & 0.68 & \textbf{0.348} & 0.371 & \textbf{0.348}$\pm$0.014 \\ \hline
\multirow{2}{*}{Ethanol} & energy  & 0.08 & 0.059 & 0.064 & \textbf{0.054} & 0.07 & - & 0.063 & 0.061$\pm$0.009\\
& forces  & 0.39 & 0.160 & 0.230 & \textbf{0.136} & 0.33 & 0.208 & 0.230 & 0.211$\pm$0.036  \\ \hline
\multirow{2}{*}{Malonaldehyde} & energy  & 0.13 & 0.094 & 0.104 & \textbf{0.081} & 0.10 & - & 0.091 & 0.096$\pm$0.013  \\
& forces  & 0.66 & 0.319 & 0.383 & \textbf{0.245} & 0.41 & 0.337 & 0.319 & 0.323$\pm$0.019 \\ \hline
\multirow{2}{*}{\textbf{Naphthalene}} & energy  & 0.16 & 0.142 & 0.122 & \textbf{0.117} & 0.12 & - & \textbf{0.117} & \textbf{0.118}$\pm$0.002 \\
& forces  & 0.58 & 0.310 & 0.215 & 0.151 & 0.11 & 0.096 & \textbf{0.083} & \textbf{0.084}$\pm$0.006\\ \hline
\multirow{2}{*}{\textbf{Salicylic Acid}} & energy  & 0.20 & 0.126 & 0.134 & 0.114 & 0.12 & - & \textbf{0.114} & \textbf{0.115}$\pm$0.008 \\
& forces  & 0.85 & 0.337 & 0.374 & 0.221 & 0.28 & 0.238 & 0.209 & \textbf{0.197}$\pm$0.004\\ \hline
\multirow{2}{*}{\textbf{Toluene}} & energy  & 0.12 & 0.100 & 0.102 & \textbf{0.098} & 0.10 & - & \textbf{0.097} & \textbf{0.094}$\pm$0.005\\
& forces  & 0.57 & 0.191 & 0.216 & 0.203 & 0.14 & 0.101 & 0.102 &  \textbf{0.088}$\pm$0.002\\ \hline
\multirow{2}{*}{Uracil} & energy  & 0.14 & 0.108 & 0.115 & \textbf{0.104} & \textbf{0.11} & - & \textbf{0.104} & \textbf{0.107}$\pm$0.004\\
& forces  & 0.56 & 0.218 & 0.301 & \textbf{0.105} & 0.24 & 0.172 & 0.140 & 0.149$\pm$0.003\\ \hline
\multirow{2}{*}{\textbf{Azobenzene}} & energy  & - & 0.197 & - & - & \textbf{0.092} & - & - & 0.142$\pm$0.003\\
& forces  & - & 0.462 & - & - & 0.409 & - & - & \textbf{0.138}$\pm$0.010\\ \hline
\multirow{2}{*}{\textbf{Paracetamol}} & energy  & - & 0.181 & - & - & 0.153 & - & - & \textbf{0.135}$\pm$0.004\\
& forces  & - & 0.519 & - & - & 0.491 & - & - & \textbf{0.263}$\pm$0.010\\
\hline\hline
    \end{tabular}
\label{tab:md17}
\end{table*}

On a similar task we train NewtonNet on the CCSD/CCSD(T) data reported for 5 small molecules\cite{Chmiela2017a, Chmiela2018}. The significance of this experiment is the gold standard of theory that is used to obtain the data, and addressing the ultimate goal to evaluate a machine learning model at high reference accuracy with an affordable number of training samples. In this benchmark data, the training and test splits are fixed at that provided by the authors of the MD17 data (i.e., 1000 training and 500 test data)\cite{Chmiela2018}. In Table \ref{tab:ccsd} we compare our results with NequIP and sGDML in which NewtonNet not only outperforms the best reported prediction performance for three of the five molecules, but it remains competitive within the range of uncertainties for the other two molecules, and is robustly improved compared to the opponent kernel methods.

\begin{table}[h]
\centering
\caption{The performance of models in terms of mean absolute error (MAE) for the prediction of energies (kcal/mol) and forces (kcal/mol/\AA) of molecules at CCSD or CCSD(T) accuracy. We randomly select 50 snapshots of the training data as the validation set and average the performance of NewtonNet over four random splits to find standard deviations. Best results in the standard deviation range are marked in bold.}
\begin{tabular}{llccc} \hline
\multicolumn{2}{l}{} & sGDML & NequIP & NewtonNet \\ \hline
\multirow{2}{*}{Aspirin} & energy  & 0.158 & -         & \textbf{0.100} $\pm$ 0.007          \\
 & forces  & 0.761 & \textbf{0.339}     & \textbf{0.356} $\pm$ 0.019          \\ \hline
\multirow{2}{*}{Benzene} & energy  & \textbf{0.003} & -         & \textbf{0.004} $\pm$ 0.001     \\
 & forces  & 0.039 & 0.018     & \textbf{0.011} $\pm$ 0.001     \\ \hline
\multirow{2}{*}{Ethanol} & energy  & \textbf{0.050} & -         &  \textbf{0.037} $\pm$ 0.019    \\
 & forces  & 0.350 & \textbf{0.217}     &  \textbf{0.236} $\pm$ 0.030    \\ \hline
\multirow{2}{*}{Malonaldehyde} & energy  & 0.248       & -     & \textbf{0.045} $\pm$ 0.004 \\
& forces  & 0.369       & 0.369 & \textbf{0.285} $\pm$ 0.038 \\ \hline
 \multirow{2}{*}{Toluene} & energy  & 0.030 & -         & \textbf{0.014} $\pm$ 0.001  \\
 & forces  & 0.210 & 0.101 & \textbf{0.080} $\pm$ 0.005 \\
        \hline\hline
    \end{tabular}
\label{tab:ccsd}
\end{table}

\textbf{Small Molecules with large Chemical Variations}. In a separate experiment to validate NewtonNet we trained it using the ANI-1 dataset to predict energies for a large and diverse set of 20 million conformations sampled from \~ 58k small molecules with up to 8 heavy atoms.\cite{smith2017ani} The challenges in regards this dataset are three-fold: first, the molecular compositions and conformations are quite diverse, with the total number of atoms ranging from 2 to 26, and with total energies spanning a range of near $3\times 10^5$ kcal/mol; second, only energy information is provided, so a well-trained network needs to extract information more efficiently from the dataset to outcompete data-intensive invariant models; finally, a machine learning model that performs well on such a diverse dataset is more transferable to unseen data, and will have a wider application domain.

\begin{table}[!htb]
\centering
\caption{The test performance of the NewtonNet model on small fractions of the original 20 million molecules ANI-1 data set of molecules with a range of sizes and conformations compared with ANNs\cite{smith2017ani}. Reported in terms of mean absolute error (MAE) for energies (kcal/mol/atom)}
\begin{tabular}{lccc}
\hline\hline
& ANI & NewtonNet & NewtonNet \\
training set size & 20,000,000 & 2,000,000 & 1,000,000 \\ \hline
energies & 1.30 & 0.65 & 0.85 \\
\hline\hline
\end{tabular}
\label{tab:ANI}
\end{table}

In Table \ref{tab:ANI} we show that by utilizing only 10\% (2M) samples of the original ANI-1 data, NewtonNet yields a MAE in energies of 0.65 kcal/mol, very near the standard definitions of chemical accuracy, and halving the error compared to ANNs using the full 20M ANI-1 dataset. Even with only 5\% of the data (1M), we achieve an MAE of 0.85 kcal/mol on energies that exceeds the original performance of the ANN network trained with all data. Note that unlike the data experiments above, atomic forces are not reported with the ANI-1 data set. Although the NewtonNet model is trained without taking advantage of additional force information for the atomic environments, it clearly confirms that the directional information are generally a significant completion to the atomic feature representation regardless of the tensor order of the output properties. 

\textbf{Methane Combustion Reaction}. The methane combustion reaction data\cite{zeng2020complex} exerts a more challenging task due to the complex nature of reactive species that are often high in energy, transient, and far from equilibrium such as free radical intermediates.  Such stress tests are important for driving \textit{ab initio} molecular dynamics simulations in which even relatively low-run DFT functionals are notoriously time-consuming and limited to small system sizes. We utilize the dataset provided by Zeng et al. \cite{zeng2020complex} that is generated through an active learning procedure and followed the same split of 13,315 snapshots to define the test set. We evaluate the performance of NewtonNet on 100\%, 10\% and 1\% of the remaining data for training and validation. 
Table \ref{tab:methane} shows that when NewtonNet is trained on all available data, it drives down the error in energies and forces significantly, resulting in an MAE of 0.50 kcal/mol/atom in energies and 1.20 kcal/mol/\AA\ in forces, thereby decreasing error in energy and force predictions by 85\% and 57\%, respectively, compared to the model in the original study.\cite{zeng2020complex} Utilizing 10\% of the data, NewtonNet has an MAE that remains close to chemical accuracy, and even using only 1\% of the data maintains superiority in performance compared to the original DeepMD model trained with the full dataset. 

\begin{table}[!htb]
\centering
\caption{The performance of NewtonNet model compared with DeepMD on 13,315 test configurations of methane combustion reaction in terms of mean absolute error (MAE) for energies (kcal/mol/atom) and forces (kcal/mol/Å). We systematically reduce the amount of training data by two orders of magnitude using NewtonNet (of which 553,997 energy and force values are available from the original data set) and compare it to the 578,731 data points used in the original paper by Zeng and co-workers\cite{zeng2020complex}}
\begin{tabular}{lcccc}
\hline\hline
& DeepMD & NewtonNet & NewtonNet&NewtonNet \\
training set size & 578,731 & 553,997 & 55,399 &  5,539\\ \hline
energies & 3.227 & 0.497 & 0.689 & 0.835\\
forces & 2.77  & 1.20 & 1.83 & 2.66\\
        \hline\hline
    \end{tabular}
\label{tab:methane}
\end{table}

\textbf{Hydrogen Combustion Reaction}. This benchmark data is newly generated for this study and probes reactive pathways of hydrogen and oxygen atoms through the combustion reaction mechanism reported by Li et. al.\cite{Li2004}, and analyzed with calculated intrinsic reaction coordinate (IRC) scans of 19 biomolecular sub-reactions from Bertels et. al. \cite{Bertels2020}. Excluding 3 reactions that are chemically trivial (diatomic dissociation or recombination reactions), we obtain configurations and energies and forces for reactant, transition, and product states for 16 out of 19 reactions. The IRC data set was further augmented with normal mode displacements and AIMD simulations to sample configurations around the reaction path. All the calculations are conducted at the $\omega$B97M-V/cc-pVTZ level of theory, and the data set comprises a total of $\sim$280,000 potential energies and $\sim$1,240,000 nuclear force vectors, and will be described in an upcoming publication.

\begin{figure*}[!htb]
\center
\includegraphics[width=0.95\textwidth]{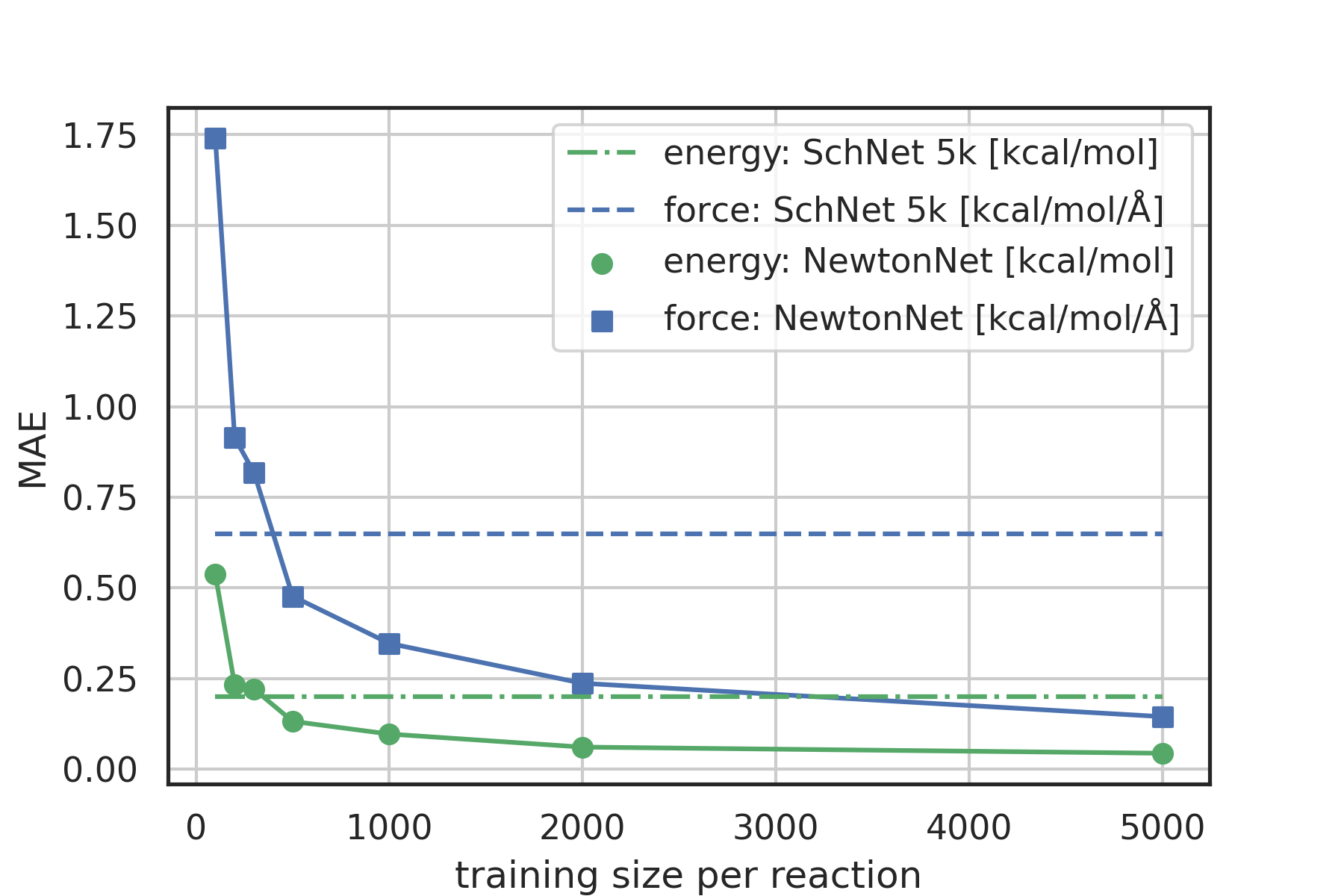}
\caption{The learning curve of NewtonNet for the hydrogen combustion data, with MAEs of energy and forces averaged over the 16 independent reactions and with respect to the number of training samples used for each reaction. The dashed lines show the performance of SchNet when trained on all 5k data per sub-reaction.}
\label{fig:hydrogen_curve}
\end{figure*}

We train NewtonNet on the complete reaction network by sampling training, validation, and test sets randomly formulated from the total data. The validation and test sizes are fixed to 1000 data per reaction, and the size of training data varies in a range of 100 to 5000 data points per reaction. The resulting model accuracy on the hold-out test set for both energy and forces is reported in Figure \ref{fig:hydrogen_curve}. It is seen that NewtonNet can outperform the best invariant SchNet model with slightly less than one order of magnitude smaller training data (500 vs 5000 samples per reaction), and is capable of achieving the chemical accuracy goal with as little as 200 data points per reaction. In conventional deep learning approaches for reactive chemistry, abrupt changes in the force magnitudes can give rise to multimodal distributions of data, which can introduce covariate shift in the training of the models. Here we posit that a better representation of atomic environments using the latent force directions can increase the amount of attention that one atom gives to its immediate neighbors. As a result the performance of NewtonNet in prediction of forces for methane and hydrogen combustion reactive systems benefit most from the directional information provided by atoms that break or form new bonds.    

\section*{\fontsize{12}{12}\selectfont DISCUSSION}


\textbf{Interpretability of Latent Force Vectors in NewtonNet}. At the core of NewtonNet message passing layers, we construct latent atomic forces to collect directional information for each atom. Here we demonstrate that the latent space in fact agrees with the ground-truth force vectors if guided by appropriate loss functions (see Methods Section). Figure \ref{fig:latent_force} shows the latent force vectors of the three consecutive message passing layers for a snapshot of the ethanol molecule from the test set of the MD17 data set. In the first layer, we may see opposite direction and orthogonality, but after the last layer transformation most of the constructed latent force vectors are in the same direction as the true reference forces, as quantified by the average cosine distance on the test set of 1.17, 0.89, and 0.05, respectively for layers one to three. The precise direction of atomic forces indicates that our model has also adequately learned the force magnitudes and signs for the pairwise interactions (based on equations \ref{eq:fij} and \ref{eq:fi}). This outcome clearly confirms that our approach in describing the latent space of the model based on the Newton's second and third law enables the model to take advantage of the underlying physics of interatomic interactions.

\begin{figure*}
\centering
\includegraphics[width=0.93\textwidth]{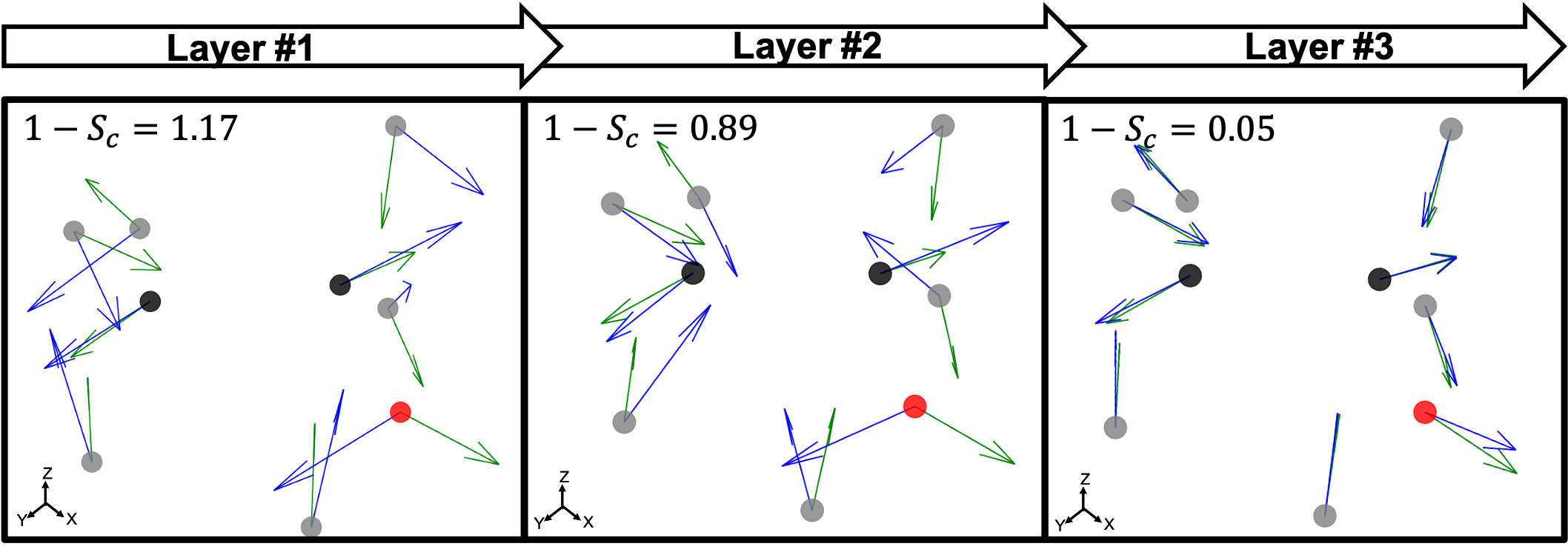}
\caption{The visualization of normalized latent force vectors (blue) and normalized reference force vectors (green) exerted on atoms of ethanol molecule from MD17 test set, represented by black:carbon, gray:hydrogen, and red:oxygen spheres. The $1-S_c \in [0,2]$ shows the average cosine distance of the two vectors for all atoms in the test set for aspirin. From left to right, figures show that cosine similarity ($S_c$) increases (cosine distance decreases) after each message passing layer of NewtonNet.}
\label{fig:latent_force}
\end{figure*}

\textbf{Computational Efficiency of NewtonNet}. In addition to data efficiency as illustrated in Results, NewtonNet allows for a linear scaling in computational complexity with respect to the number of atoms in the system. To give a better sense of the computational efficiency we compare the time that is needed to train on the aspirin molecule from the MD17 data set with the same calculation using the NequIP model. As reported by Batzner et. al., a complete training on the MD17 data to converge to the best performance of NequIP model takes up to 8 days.\cite{batzner21} However, NewtonNet only required 12 hours to give the state-of-the-art performance on a GeForce RTX 2080Ti, GPU which is only 73\% as fast as the Tesla V100 that is used for evaluating NequIP, when a straightforward comparison of similar rank of contributed tensors used by both methods. Obviously, higher order tensors may boost performance but will increase the computation time, and should be analyzed from a cost-benefit perspective to find the best level of ML models for the required accuracy versus computational resources. 

Aside from training time that is important to facilitate the model development and to reduce the testing time, the computation time per atomic environment is critical for the future application of trained models in an MD simulation. The computation time for processing a snapshot of the MD trajectory of a small molecule by NewtonNet is 4 milliseconds ($\sim 3$ ms on a Tesla V100) for a small molecule of 20 atoms. Considering the reported average time of 16 milliseconds for NequIP to process a molecule of 15 atoms\cite{batzner21}, NewtonNet demonstrates a significant speedup. In addition, the PaiNN model\cite{Schutt2021} is the closest to our model in terms of computational complexity, but does not encode additional physical knowledge in the message passing operators. As a result, it includes about 20\% more number of optimized parameters (600k vs. 500k parameters in NewtonNet). This difference likely leads to higher computational cost with an equally efficient implementation of the code. Nevertheless, all these reported prediction times are by far smaller than \textit{ab initio} calculations even for a snapshot of a small molecule in the MD trajectory, which is on the order of minutes to hours. 

\section*{\fontsize{12}{12}\selectfont CONCLUSIONS}
\label{sec:conclusion}
\noindent
The ability to predict the energy and forces of a molecular dynamics trajectory with high accuracy but at an efficient time scale is of considerable importance in the study of chemical and biochemical systems. We have developed a new ML model based on Newton's equation of motion that can conduct this task more accurately (or achieve competitive performance) than other state-of-the-art invariant and equivariant models. 

Overall, the presented results are promising in at least three major respects. First, since the NewtonNet model takes advantage of geometric message passing and a rotationally equivariant latent space which scales linearly with the size of the system, its promising performance in accuracy can be achieved without much computation or memory overhead. NewtonNet, like other equivariant deep learning models, utilizes less data and can still outperform the kernel methods that are renowned for their good performance on small size data. Given the better scalability of deep learning models such as NewtonNet compared to kernel methods, we can expand the training data, for example by smart sampling methods like active learning\cite{haghighatlari2020, Miksch2021} and explore the potential energy surface of the chemical compound space more efficiently. The study of methane combustion reaction is a proof of evidence for this approach as the training data is a result of active learning sampling. If this sampling was initiated with NewtonNet predictions, one could achieve the best performance with even less number of queries. 

Second, the data efficiency is the key to achieve ML force field models at the high accuracy levels of \emph{first principles} methods such as CCSD(T)/CBS with competitive performances as state-of-the-art kernel-based methods using significantly less training data. For the CCSD(T) data on small single organic molecules we found that the NewtonNet performance is competitive or better than state-of-the-art equivariant models by at least 10\%. This is a very encouraging result for being able to obtain gold standard levels of theory with affordable data set generation. 

Finally, taking advantage of Newton's laws of motion in the design of the architecture helped to avoid unnecessary operations, and provide a more understandable and interpretable latent space to carry out the final predictions. Inspired by other physical operations that incorporate higher order tensors\cite{batzner21,Qiao2021}, NewtonNet can also be further extended to construct more distinguishable latent space many-body features in future work. Even so, the performance of the NewtonNet model on the MD trajectories from combustion reactions are both excellent with good chemical accuracy even when considering the challenge of chemical reactivity. 

\section*{\fontsize{12}{12}\selectfont APPENDIX A}
\textbf{Proof of Equivariance and Invariance}. We prove that our model is rotationally equivariant on the atomic positions $\mathbb{R}_i \in \mathbb{R}^\mathrm{3}$ and atomic numbers $Z_i$ for a rotation matrix $ T\in\mathbb{R}^\mathrm{3\times3}$. 
In equation 5, the euclidean distance is invariant to the rotation, as it can be shown that

\begin{align}
    \begin{split}
        \left\|T \mathbf{r}_{ij}\right\|^{2} = \\
        \left\|T \mathbf{R}_{j}-T \mathbf{R}_{i}\right\|^{2}=\\
        \left(\mathbf{R}_{j}-\mathbf{R}_{i}\right)^{\top} T^{\top} T\left(\mathbf{R}_{j}-\mathbf{R}_{i}\right)=\\
        \left(\mathbf{R}_{j}-\mathbf{R}_{i}\right)^{\top} \mathbf{I}\left(\mathbf{R}_{j}-\mathbf{R}_{i}\right)=\\
        \left\|\mathbf{R}_{j}-\mathbf{R}_{i}\right\|^{2}=\\
        \left\| \mathbf{r}_{ij}\right\|^{2},
    \end{split}
\end{align}
\noindent
which means that the Euclidean distance is indifferent to the rotation of the positions as it is quite well-known for this feature. Consequently, feature arrays $m_{ij}$, $a_i$, and all the linear or non-linear functions acting on them will result in invariant outputs. The only assumptions for this proof is that a linear combination of vectors or their product with invariant features will remain rotationally equivariant. Based on this assumption we claim that equations \ref{eq:fij} to \ref{eq:dr_ext} will remain equivariant to the rotations. For instance, the same rotation matrix $T$ propagates to equation \ref{eq:fij} such that,

\begin{align}
    \left.\phi_{F}(T{m}_{i j}\right.) T\hat{r}_{i j} = \left.\phi_{F}({m}_{i j}\right.) T\hat{r}_{i j}= T \left. \phi_{F}({m}_{i j}\right.) \hat{r}_{i j}=  T\vv{F}_{ij}^{l}.
\end{align}

The last operator, equation \ref{eq:du}, will remain invariant to the rotations due to the use of dot product. The proof for the invariant atomic energy changes is that,

\begin{align}
    \begin{split}
         - \phi_{u}({a}_{i}^{l+1}) \left. (T\bm{f}_{i}^{l+1} \cdot T\bm{dr}_{i}^{l+1}\right.) = \\
         - \phi_{u}({a}_{i}^{l+1}) \left. (\bm{f}_{i}^{l+1} \ T^{\top} T \ \bm{dr}_{i}^{l+1}\right.) = \\
         - \phi_{u}({a}_{i}^{l+1}) \left. (\bm{f}_{i}^{l+1} \ \mathbf{I} \ \bm{dr}_{i}^{l+1}\right.) = \\
         - \phi_{u}({a}_{i}^{l+1}) \left. (\bm{f}_{i}^{l+1} \cdot \bm{dr}_{i}^{l+1}\right.) = \\
         {\delta U}_{i}.
    \end{split}
\end{align}
\noindent
This is how we contract equivariant features to invariant arrays. The addition of these arrays to atomic features preserves the invariance for the final prediction of total potential energy based on atomic contributions.

\section*{\fontsize{12}{12}\selectfont METHODS}
\label{sec:methods}
\textbf{Training Details}. We follow the summation rule as described by Behler and Parrinello\cite{Behler2007} to predict the atomic energies. Following this rule, we use a differentiable function to map the updated atomic features after last layer $a_i^L$ to atomic potential energies $E_i$. Ultimately, the total potential energy is predicted as the sum of all atomic energies.

\begin{align}
    {E}_{i} = \phi_{out} ({a}_{i}^{L}), \\
    \widetilde{E} = \sum_{i}^{N_m} {E}_{i},
\end{align}
where $N_m$ is the total number of atoms, and $\phi_{out}: \mathbb{R}^{\mathrm{nf}} \rightarrow \mathbb{R}^\mathrm{1}$ is a fully connected network with Sigmoid Linear Unit (SiLU) activation\cite{Elfwing2018} after each layer except the last layer.

We obtain forces as gradient of potential energy with respect to atomic positions. This way we guarantee the energy conservation\cite{Christensen2020} and provide atomic forces for a robust training of the atomic environments,

\begin{align}
    \bm{\widetilde{F_i}} = -\nabla_i \widetilde{E}.
\end{align}

We train the model using small batches of data with batch size $M$. The loss function penalizes the model for predicted energy values, force components, and the direction of latent force vectors from last message passing layer $\bm{\mathcal{F}}_i^L$. These three terms of the loss function $\mathcal{L}$ are formulated as: 

\begin{align}
    \mathcal{L}=\frac{\lambda_{E}}{M} \Sigma_{m}^{M}\left(\widetilde{E}_{m}-E_{m}\right)^{2}+\\ \nonumber
    \frac{\lambda_{F}}{M} \Sigma_{m}^{M} \frac{1}{3 N_{m}} \sum_{i}^{N_{m}}\left\|\widetilde{\bm{F}}_{m i}-\bm{F}_{m i}\right\|^{2}+\\ \nonumber
    \frac{\lambda_{D}}{M \times N_{m}} \sum_{m}^{M} \sum_{i}^{N_{m}}\left(1-\frac{{\bm{\mathcal{F}}}_{m i}^{L} \cdot \bm{F}_{m i}}{\left\|{\bm{\mathcal{F}}}_{m i}^{L}\right\|\left\|\bm{F}_{m i}\right\|}\right).
\end{align}

The first two terms are common choices for the energy and forces that are on the basis of the mean squared deviations of predicted values with references data. The last term penalizes the deviation of latent force vectors direction with the ground-truth force vectors. Here, we use cosine similarity loss function to minimize the $(1- \cos(\alpha)) \in [0,2]$, where $\alpha$ is the angle between the $\bm{\mathcal{F}}^L_{i}$ and $\bm{F_{i}}$ for each atom $i$ of a snapshot $m$ of a MD trajectory. The $\lambda_E$, $\lambda_F$, and $\lambda_D$ are hyperparameters that determine the contribution of energy, force, and latent force direction losses in the total loss $\mathcal{L}$.

\begin{table}[!htb]
\centering
\caption{Hyperparameters for all the reported experiments in the results section.}
\begin{tabular}{lcccccc}
\hline\hline
& $\lambda_E$ & $\lambda_F$ & $\lambda_D$ & learning rate (lr) & lr decay & cutoff radius [\AA]\\\hline
MD17            & 1 & 50 & 1  &  $1.10^{-3}$ & 0.7 & 5\\
MD17/CCSD(T)    & 1 & 50 & 1  & $1.10^{-3}$ & 0.7 & 5\\
ANI             & 1  & 0 & 0 & $1.10^{-4}$ & 0.7 & 5\\
Methane combustion             & 1  & 30 & 1 & $1.10^{-3}$ & 0.7 & 5\\
Hydrogen combustion             & 1  & 20 & 1 & $5.10^{-4}$ & 0.7 & 5\\

        \hline\hline
    \end{tabular}
\label{tab:app_hyperparam}
\end{table}

We use mini-batch gradient descent algorithm (with Adam optimizer \cite{Kingma2015}) to minimize the loss function with respect to the trainable parameters. The trainable parameters are built in the learned functions noted with $\phi$ symbol. We use fully connected neural network with SiLU nonlinearity for all $\phi$ functions through out the message passing layer. The only exception is the $\phi_{rbf}$, which is a single linear layer. We avoid using bias parameters in the $\phi_{f}$ and $\phi_{r}^{'}$ in order to propagate the radial cutoff throughout the network. We found it important for the ANI model to use a normalization layer\cite{Ba2015} on the atomic features at every message passing layer as it helps with the stability of training. All NewtonNet models in this paper use $L=3$ message passing layers, $nf=128$ features and $nb=20$ basis sets. The number of features are set similar to previous works to emphasize on the impact of architecture design in our comparisons. Other hyper-parameters are selected based on the best practices for each type of system and are reported in the Table \ref{tab:app_hyperparam}.

For the training of SchNet in the hydrogen combustion study we use 128 features everywhere and 5 interaction layers as recommended by developers\cite{schutt2018schnetpack}. The other hyperparameters are the same as NewtonNet except for the force coefficient in the loss function that we found a lower $\lambda_F = 10$ performs better than larger coefficients.

\section*{ACKNOWLEDGMENTS}
\noindent
The authors thank the National Institutes of Health for support under Grant No 5U01GM121667. FHZ thanks the Research Foundation-Flanders (FWO) Postdoctoral Fellowship for support as a visiting Berkeley scholar. M. Liu thanks the China Scholarship Council for a visiting scholar fellowship. C.J.S. acknowledges funding by the Ministry of Innovation, Science and Research of North Rhine-Westphalia (“NRW Rückkehrerprogramm”) and an Early Postdoc.Mobility fellowship from the Swiss National Science Foundation. This research used computational resources of the National Energy Research Scientific Computing Center, a DOE Office of Science User Facility supported by the Office of Science of the U.S. Department of Energy under Contract No. DE-AC02-05CH11231.
\newline
\section*{AUTHOR CONTRIBUTIONS}
\noindent
T.H-G. and M.H. conceived the scientific direction and wrote the complete manuscript. M.H., J.L., X.G., O.Z., M.L., and H.H. ran experiments with NewtonNet and other ML models on the various datasets and wrote the corresponding Results in the manuscript. A.D., C.J.S., L.B. and F.H-Z. generated the hydrogen combustion data. T.H-G., M.H., J.L., O.Z., F.H-Z., and X.G analyzed the results. All authors provided comments on the results and manuscript.
\newline
\section*{DECLARATION OF INTERESTS}
\noindent
The authors declare no competing interests.
\newline
\bibliographystyle{unsrtnat}
\bibliography{references}

\end{document}


\maketitle

\section*{\fontsize{12}{12}\selectfont NewtonNet Modules}

The core idea behind the iterative message passing of the atomic environments is to update the feature array ${a}_{i}^{t}$ that represent each atom in its immediate environment. This level of feature representation is invariant to the rotations of the initial configuration in NewtonNet through the following operations.

\textbf{Atomic Feature Aggregator.} We initialize the atomic features based on trainable embedding of atomic numbers $Z_i$, i.e., ${a}_{i}^{0} = g(Z_i)$ and ${g}: \mathbb{R}^{\mathrm{1}} \rightarrow \mathbb{R}^\mathrm{nf}$. We next use the edge function ${e}: \mathbb{R}^{\mathrm{3}} \rightarrow \mathbb{R}^\mathrm{nb}$ to represent the interatomic distances using radial Bessel functions as introduced by Klicpera et al.\cite{?}

\begin{align}
    e(\vv{r_{i j}})=\sqrt{\frac{2}{r_{c}}} \frac{\sin (\frac{n \pi}{r_{c}} \lVert \vv{r_{i j}} \rVert )}{\lVert \vv{r_{i j}} \rVert}
\end{align}

where $r_c$ is the cutoff radius and $\lVert \vv{r_{i j}} \rVert$ returns the interatomic distance between any atom $i$ and $j$.
We follow Schutt et al.\cite{?} in using a self-interaction linear layer $\phi_{rbf}: \mathbb{R}^{\mathrm{nb}} \rightarrow \mathbb{R}^\mathrm{nf}$ to combine the output of radial basis functions with each other. This operation is followed by using an envelop function to implement a continuous radial cutoff around each atom. For this purpose, we use the polynomial function $e_{cut}$ introduced by Klicpera et al.\cite{?} with the choice of degree of polynomial $p=7$. Thus, the edge operation ${\phi_e}: \mathbb{R}^{\mathrm{3}} \rightarrow \mathbb{R}^\mathrm{nf}$ is defined as a trainable transformation of relative atom position vectors in the cutoff radius ${r}_{c}$

\begin{align}
    \phi_{e}(\vv{r_{i j}}) = \phi_{rbf}(e(\vv{r_{i j}})) \left e_{cut}(r_c, \lVert \vv{r_{i j}} \rVert\right).
\end{align}

The output of $\phi_{e}$ is rotationally invariant as it only depends on the interatomic distances.
Following the notation of neural message passing, we define a message function to collect the neighboring information and update atomic features. Here, we tend to pass a symmetric message between any pair of atoms, i.e., the message that is passed between atom $i$ and atom $j$ are the same in both directions. Thus, we introduce our symmetric message passing $m_{ij}$ by element-wise product between all feature arrays involved in any two-body interaction,

\begin{align}
    {m}_{i j} = \phi_{a}({a}_{i}^{l}) \left \phi_{a}({a}_{j}^{l})\right \phi_{e}(\vv{r_{i j}})
\end{align}

where $\phi_{a}: \mathbb{R}^{\mathrm{nf}} \rightarrow \mathbb{R}^\mathrm{nf}$ indicates a trainable and differentiable network with a nonlinear activation function SiLU \cite{?} after the first layer.
Note that the $\phi_{a}$ is the same function applied to all atoms. Thus, due to the weight sharing and multiplication of output features of both heads of the two-body interaction, the ${m}_{i j}$ remain symmetric at each layer of message passing. To complete the feature array aggregator, we use the equation \ref{eqn:aggregate} to simply sum all messages received by central atom $i$ from its neighbors $\mathcal{N}(i)$. Finally, we update the atomic features at each layer using the sum of received messages,
\begin{align}
    {a}_{i}^{l+1} = {a}_{i}^{l} + \sum_{j \in \mathcal{N}(i)} {m}_{ij}.
\end{align}

\emph{\textbf{force calculator.}} So far, we have followed a standard message passing that is invariant to the rotation. We begin to take advantage of directional information starting from the force calculator module. The core idea behind this module is to construct latent force vectors using the Newton's third law. The third law states that the force that atom $i$ exerts on atom $j$ is equal and in opposite direction of the force that atom $j$ exerts on atom $i$. This is the reason that we intended to introduce a symmetric message passing operator. Thus, we can estimate the symmetric force magnitude as a function of ${m}_{i j}$, i.e., $\lVert \vv{F}_{ij}\rVert = \phi_{F}({m}_{i j})$. The product of the force magnitude by unit distance vectors $\hat{r}_{i j} = \vv{r}_{i j}/\lVert \vv{r}_{i j} \rVert$ gives us antisymmetric interatomic forces that obey the Newton's third law (note that $\vv{r}_{i j} = -\vv{r}_{j i}$),

\begin{align}
    \vv{F}_{ij}^{l} = \left\phi_{F}({m}_{i j}\right) \hat{r}_{i j}
\end{align}
where $\phi_{F}: \mathbb{R}^{\mathrm{nf}} \rightarrow \mathbb{R}^\mathrm{1}$ is a differentiable learned function, and $\vv{F}_{ij}^{l} \in \mathbb{R}^{\mathrm{3}}$. The total force at each layer $\vv{F}_{i}^{l}$ on atom $i$ is the sum of all the forces from the neighboring atoms $j$ in the atomic environment,  
\begin{align}
    \vv{F}_{i}^{l} = \sum_{j \in \mathcal{N}(i)} \vv{F}_{ij}^{l},
\end{align}
and updating the latent force vectors at each layer,
\begin{align}
    \bm{\mathcal{F}}_{i}^{l+1} = \bm{\mathcal{F}}_{i}^{l} + \vv{F}_{i}^{l}.
\end{align}

We ultimately use the latent force vector from the last layer $L$, $\bm{\mathcal{F}}_{i}^{L} \in \mathbb{R}^{\mathrm{3}}$ in the loss function to ensure this latent space truly mimics the underlying physical rules. 

To complete the force calculator module, we borrow the idea of continuous filter from Schut et al.\cite{} to decompose and scale latent force vectors along each dimension using another learned function $\phi_{f}: \mathbb{R}^{\mathrm{nf}} \rightarrow \mathbb{R}^\mathrm{nf}$. This way we can featurize the vector field to avoid too much of abstraction in the structural information that they carry with themselves,
\begin{align}
    \bm{\Delta f}_{i} = \sum_{j \in \mathcal{N}(i)} \left\phi_{f}({m}_{ij}\right) \vv{F}_{ij}^{l}.
\end{align}
As a result, the constructed latent interatomic forces are decomposed by rotationally invariant features along each dimmension, i.e., $\bm{\Delta f}_{i} \in \mathbb{R}^{\mathrm{3 \times nf}}$. We call this type of representation feature vectors. Following the message passing strategy, we update the force feature vectors with $\bm{\Delta f}_{i}$ after each layer, while they are initialized with zero values, $\bm{f}_{i}^{0} = \bm{0}$,
\begin{align}
    \bm{f}_{i}^{l+1} = \bm{f}_{i}^{l} + \bm{\Delta f}_{i}.
\end{align}

\emph{\textbf{momentum calculator.}} This is the step that we try to estimate a measure of atomic displacement due to the forces that are exerted on them. We accumulate their dispalcements at each layer without updating the position of each atom. The main idea in this module is that the displacement must be along the updated force features in the previous step. Inspired by Newton's second law, we approximate the displacement factor using a learned function $\phi_{r}: \mathbb{R}^{\mathrm{nf}} \rightarrow \mathbb{R}^\mathrm{nf}$ that acts on the current state of each atom presented by its atomic features ${a}_{i}^{l}$,
\begin{align}
    \bm{\delta r}_{i} = \phi_{r}({a}_{i}^{l+1}) \bm{{f}_{i}^{l+1}}.
\end{align}
We finally update the displacement feature vectors by $\bm{\delta r}_{i}$ and a weighted sum of all the atomic displacements from the previous layer. The weights are estimated based on a trainable function of messages ($\phi_{r}^{'}: \mathbb{R}^{\mathrm{nf}} \rightarrow \mathbb{R}^\mathrm{nf}$) between atoms,
\begin{align}
    \bm{dr}_{i}^{l+1} = \sum_{j \in \mathcal{N}(i)} \phi_{r}^{'} \left({m}_{i j}\right) \bm{dr}_{i}^{l} + \bm{\delta r}_{i}. 
\end{align}
The weight component in this step works like attention mechanism to concentrate on the two-body interactions that cause maximum movement in the atoms. Since forces at $l=0$ are zero, the displacements are also initialized with zero values, i.e., $\bm{{dr}_{i}^{0}} = \bm{0}$.

\emph{\textbf{energy calculator.}} The last module contracts the directional information to the rotationally invariant atomic features. Since we developed the previous steps based on the Newton's equations of motion, one immediate idea is to approximate the potential energy change for each atom using $f_i^l$ and $\delta r_i^l$, considering that $\bm{f_i^l} \approx - \delta U / \bm{\delta r_i^l}$. Thus, we find energy change for each atom by

\begin{align}
    {\delta U}_{i} = - \phi_{u}({a}_{i}^{l+1}) \left\langle \bm{f}_{i}^{l+1} \cdot \bm{dr}_{i}^{l+1}\right\rangle,
\end{align}
where ${\delta U}_{i} \in \mathbb{R}^\mathrm{nf}$ and $\phi_{u}: \mathbb{R}^{\mathrm{nf}} \rightarrow \mathbb{R}^\mathrm{nf}$ is a differentiable learned function that operates on the atomic features and predicts the energy coefficient for each atom. The dot product of two feature vectors contracts the features along each dimension to a single feature array. We finally update the atomic features once again using the contracted directional information presented as atomic potential energy change,

\begin{align}
    {a}_{i}^{l+1} = {a}_{i}^{l+1} + {\delta U}_{i}.
\end{align}
This approach is both physically and mathematically consistent with the rotational equivariance operations and the goals of our model development. Physically, the energy change is the meaningful addition to the atomic feature arrays as they are used to predict the atomic energies eventually. Mathematically, the dot product of two feature vectors contracts the rotationally equivariant features to invariant features similar to euclidean distance that we used in the \emph{atomic feature aggregator} module.

\section*{\fontsize{12}{12}\selectfont Proof of Equivariance and Invariance}

We prove that our model is rotationally equivariant on the atomic positions $\mathbb{R}_i \in \mathbb{R}^\mathrm{3}$ and atomic numbers $Z_i$ for a rotation matrix $ T\in\mathbb{R}^\mathrm{3\times3}$. 
In the equation 1, the euclidean distance is invariant to the rotation, as it can be shown that

\begin{align}
    \begin{split}
        \left\|T \mathbf{r}_{ij}\right\|^{2} = \\
        \left\|T \mathbf{R}_{j}-T \mathbf{R}_{i}\right\|^{2}=\\
        \left(\mathbf{R}_{j}-\mathbf{R}_{i}\right)^{\top} T^{\top} T\left(\mathbf{R}_{j}-\mathbf{R}_{i}\right)=\\
        \left(\mathbf{R}_{j}-\mathbf{R}_{i}\right)^{\top} \mathbf{I}\left(\mathbf{R}_{j}-\mathbf{R}_{i}\right)=\\
        \left\|\mathbf{R}_{j}-\mathbf{R}_{i}\right\|^{2}=\\
        \left\| \mathbf{r}_{ij}\right\|^{2},
    \end{split}
\end{align}

which means that the euclidean distance is indifferent to the rotation of the positions as it is quite well-known for this feature. Consequently, feature arrays $m_{ij}$, $a_i$, and all the linear or non-linear functions acting on them will result in invariant outputs. 
The only assumptions for this proof is that a linear combination of vectors or their product with invariant features will remain rotationally equivariant. Base on this assumption we claim that equation 5 to 11 will remain equivariant to the rotations. For instance, the same rotation matrix $T$ propagates to equation 5 such that,

\begin{align}
    \left\phi_{F}(T{m}_{i j}\right) T\hat{r}_{i j} = \left\phi_{F}({m}_{i j}\right) T\hat{r}_{i j}= T \left \phi_{F}({m}_{i j}\right) \hat{r}_{i j}=  T\vv{F}_{ij}^{l}.
\end{align}

The last operator, equation 12, will remain invariant to the rotations due to the use of dot product. The proof for the invariant atomic energy changes is that,

\begin{align}
    \begin{split}
         - \phi_{u}({a}_{i}^{l+1}) \left (T\bm{f}_{i}^{l+1} \cdot T\bm{dr}_{i}^{l+1}\right) = \\
         - \phi_{u}({a}_{i}^{l+1}) \left (\bm{f}_{i}^{l+1} \ T^{\top} T \ \bm{dr}_{i}^{l+1}\right) = \\
         - \phi_{u}({a}_{i}^{l+1}) \left (\bm{f}_{i}^{l+1} \ \mathbf{I} \ \bm{dr}_{i}^{l+1}\right) = \\
         - \phi_{u}({a}_{i}^{l+1}) \left (\bm{f}_{i}^{l+1} \cdot \bm{dr}_{i}^{l+1}\right) = \\
         {\delta U}_{i}.
    \end{split}
\end{align}

This is how we contract equivariant features to invariant arrays. The addition of these arrays to atomic features preserves the invariance for the final prediction of atomic contributions to the total potential energy.

\bibliographystyle{chem-acs}
\bibliography{references}